# Vorstoss und Rückgang der Gletscher während der Kleinen Eiszeit

Matthias Huss[1,2] and Simon Förster[1]

[1]Laboratory of Hydraulics, Hydrology and Glaciology (VAW), ETH Zurich, Zurich, Switzerland
[2]Department of Geosciences, University of Fribourg, Fribourg, Switzerland

## 1. Einleitung

Die Gletscher in den Alpen sind wichtige und eindeutige Indikatoren für die Veränderung des Klimas. Seit ihrem letzten Hochstand um 1850 ziehen sie sich kontinuierlich zurück und haben seither rund zwei Drittel ihres Volumens eingebüsst (Zemp et al., 2006). Während der sogenannten «Kleinen Eiszeit» zwischen etwa 1350 und 1850 hingegen war das Klima in der Schweiz deutlich kühler (Denton and Karlén, 1973; Casty et al., 2005) und die Gletscher verzeichneten verschiedene Vorstoss-Phasen, die zu den heute im Gelände sichtbaren Moränenwällen führten. Diese markieren die ehemalige Ausdehnung der Gletscher eindrücklich. Während der letzten 10'000 Jahren waren die Alpen-Gletscher nie mehr ähnlich gross (Holzhauser et al., 2005). Die Vorstösse der Gletscher, teilweise bis weit in die Täler, führte zu verschiedenen Problemen: Einerseits wurde wichtiges Weideland zerstört und die ungünstigen Klimabedingungen führten zu Ernteausfällen und Hungersnöten, die unter anderem am Ursprung der Französischen Revolution standen (Fagan, 2000). Andererseits sind verschiedene Fälle dokumentiert, wo verstossende Gletscherzungen Seen aufstauten, welche dann zu fatalen Fluten führten. Solche zerstörerischen Seeausbrüche während der Kleinen Eiszeit sind nicht nur für den Glacier du Giétroz im Val de Bagnes für 1595 und 1818 bekannt, sondern sind auch für den Allalingletscher im Saastal in den Jahren 1589, 1633, 1680 und 1772, sowie den Vernagtferner im Ötztal, Österreich, um 1600, 1678 und 1845-1848 beschrieben (Rabot, 1905; VAW, 2003).

Obwohl die Gletschervorstösse in den Alpen um 1850 sehr gut dokumentiert sind, steht der Grund für den Zeitpunkt dieser maximalen Ausdehnung, sowie ihren anschliessenden, sehr schnellen Rückgang noch immer unter wissenschaftlicher Debatte. Tatsächlich zeigen Messungen der Temperatur nämlich, dass kühle Bedingungen noch bis in die 1910er Jahre vorherrschten (Casty et al., 2005; Auer et al., 2006). Man würde also erwarten, dass die Gletscher ihren Maximalstand nicht um 1850, sondern deutlich später erreichten, was hingegen den langen Messreihen klar widerspricht (Holzhauser&Zumbühl, 1999). Das Problem der massiven Gletschervorstösse und dem anschliessenden Rückgang wird deshalb oft als das «Paradox der Kleinen Eiszeit» bezeichnet (Vincent et al., 2005).

Da die Lufttemperatur nicht der einzige Grund für die starken Gletscher-Veränderungen in der letzten Phase der Kleinen Eiszeit sein kann, wurden verschiedene andere Erklärungs-Ansätze vorgeschlagen. Vincent et al. (2005) machen einen Anstieg der Niederschläge für die Gletschervorstösse verantwortlich. Ein solcher Anstieg ist aber weder in den wenigen direkten Messungen in dieser Zeit, noch in den Klima-Rekonstruktionen abgebildet. Nussbaumer et al. (2011) und Lüthi (2014) gehen davon aus, dass reduzierte Sonnen-Einstrahlung (geringere Aktivität der Sonne, häufige Vulkanausbrüche) einen wichtigen Einfluss auf die Gletschervorstösse ausgeübt haben könnte, konnten dies aber nicht quantitativ belegen. Painter et al. (2013) stellten die Theorie auf, dass der schnelle Rückgang der Gletscher nach 1850 auf die verstärkte Ablagerung von Verbrennungsrückständen (sogenanntes «Black Carbon») auf Gletschern zurückzuführen ist. Die erhöhte Konzentration der Russpartikel wird dabei dem Beginn der Industrialisierung zugeschrieben und beschleunigt die Gletscher-Schmelze dank einer Reduktion der Rückstrahlfähigkeit von Schnee und Eis. Allerdings konnte eine kürzlich erschienene Studie zeigen, dass dieser Effekt wahrscheinlich nur gering war und erst einige Jahrzehnte nach dem beobachteten Einsetzen des Gletscherschwunds aufgetreten ist (Sigl et al., 2018).

Hier möchten wir die Veränderung der Gletscher in der Endphase der Kleinen Eiszeit mit einer Zusammenstellung verschiedener Datensätze und einer einfachen Analyse betrachten, um daraus Schlussfolgerungen zu den Gründen der starken Vorstösse, sowie des schnellen Rückgang der Gletscherzungen nach 1850 zu finden. Damit versuchen wir auch eine direkte Verbindung zwischen den Prozessen, die der Gletscher-Veränderung zugrunde liegen, und dem katastrophalen Gletschersee-Ausbruch im Juni 1818 im Val de Bagnes aufzuzeigen.

## 2. Daten

Gletscherrückgang und -vorstoss ist für verschiedene Gletscher in den Alpen über mehrere Jahrhunderte dokumentiert. Seit rund 1880 werden im Rahmen des Schweizer Gletschermessnetzes (GLAMOS) jährlich detaillierte Messungen an den Gletscherzungen durchgeführt. Vor dieser Zeit konnte durch eine umfangreiche Interpretation von historischen Gemälden verschiedener Künstler eine genaue Rekonstruktion erstellt werden (Holzhauser&Zumbühl, 1999; Holzhauser et al., 2005; Zumbühl et al., 2008; Nussbaumer et al., 2011). Die Daten zeigen, dass sich die Gletscher zwischen dem 16$^{ten}$ und dem frühen 19$^{ten}$ Jahrhundert verschiedenen Vorstoss-Phasen hatten und fast so gross wie während ihrer maximalen Ausdehnung um 1850 waren (Abb. 1). Danach setzte ein schneller Gletscher-Rückgang ein, der bis heute mit einigen kurzen Unterbrüchen – letztmals in den 1980er Jahren – anhält.

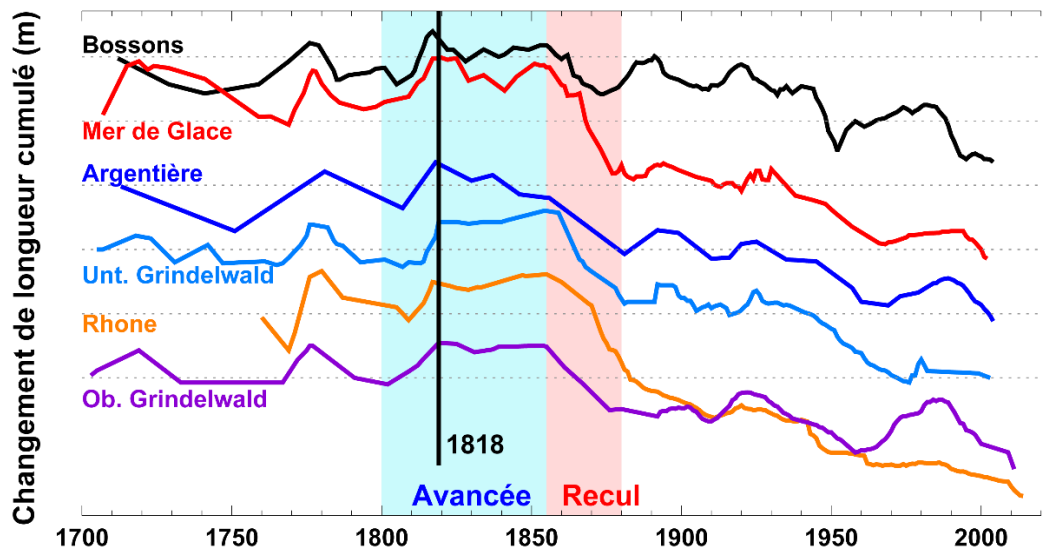

**Abb. 1 : Beobachteter Vorstoss und Rückgang von sechs Gletschern in den Alpen seit 1700 (Holzhauser&Zumbühl, 1999; Zumbühl et al., 2008). Die Vorstossperiode der Gletscher zwischen 1800 und 1850, sowie ihr anschliessender, markanter Rückgang ist gekennzeichnet. Der Zeitpunkt des Seeausbruchs am Glacier du Giétroz im Jahr 1818 ist hervorgehoben.**

Zur Interpretation der Gletscher-Veränderungen verwenden wir die direkten Messungen von Lufttemperatur und Niederschlag aus dem HISTALP Datensatz (Auer et al., 2006). Die längsten Messreihen im Westen der Alpen reichen für die Temperatur bis ins Jahr 1760, für den Niederschlag bis 1800 zurück.

Für die Höhenlagen der Gletscher gibt es allerdings erst später gesicherte Informationen. Die Datenreihen zeigen, dass die tiefsten Sommer-Temperaturen, die relevanteste Grösse für die Gletscher-Schmelze, um rund 1910 erreicht wurden (Abb. 2). Die Sommer waren damals also im Schnitt noch kälter als während der Gletscher-Vorstösse in der ersten Hälfte des 19ten Jahrhunderts. Im Vergleich dazu sind die Sommer-Temperaturen bis heute um rund 2 Grad angestiegen, während sich die Niederschläge nicht wesentlich verändert haben (Abb. 2). Auffällig ist aber auch das kälteste, einzelne Jahr 1816, das als das «Jahr ohne Sommer» in die Geschichtsbücher einging (Luterbacher&Pfister, 2015).

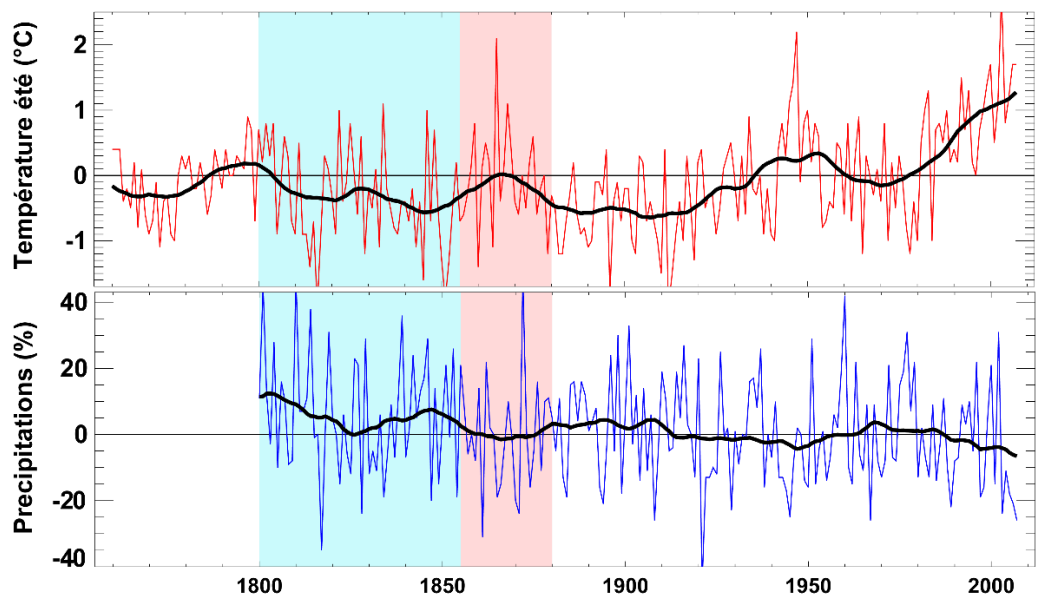

**Abb. 2 : Zeitreihen der längsten Messungen von Lufttemperatur im Sommer und Jahresniederschlag als Abweichung vom Mittel der Periode 1901-2000 für den südwestlichen Teil der Alpen (Auer et al., 2006). Die schwarzen Linien zeigen den langfristigen Verlauf. Die Vorstossperiode der Gletscher zwischen 1800 und 1850, sowie ihr anschliessender, markanter Rückgang ist gekennzeichnet.**

Zusätzlich arbeiten wir mit einem Datensatz, der alle grösseren vulkanischen Eruptionen der letzten Jahrhunderte nach geschichtlichen Quellen beschreibt (NGDC/WDS, 2018). Die Stärke des Ausbruchs wird mit einem Index quantifiziert. Wir nehmen an, dass sehr starke Vulkanausbrüche eine grosse Menge an Schwefeldioxid bis in die Stratosphäre bringen. Dort bilden sich sogenannte Sulfat-Aerosole, welche sich über den ganzen Planten verbreiten und damit massgeblich zur Reduktion der Sonneneinstrahlung beitragen.

## 3. Methoden

Wir analysieren die Veränderung des Gletschervolumens über die Zeitperiode 1760 bis 2008 mit einem stark vereinfachten Berechnungs-Modell. Dieses beschreibt die wichtigsten Grössen, welche zum Massengewinn/-verlust, bzw. einem Vorstoss/Rückgang der Gletscher beitragen. Der Einfluss der Variablen Temperatur und Niederschlag auf die Massenbilanz des Gletschers wird mit sogenannten Sensitivitäten beschrieben (Oerlemans&Reichert, 2000). So berechnete Abweichungen der Gletscher-Massenbilanz im einzelnen Jahr werden anschliessend auf die Gletscherfläche der ganzen Schweiz hochgerechnet und mit den aus anderen Datensätzen bekannten Verlusten des Eisvolumens in der Periode 1900 bis 2008 verglichen (Huss, 2012).

Es ist bekannt, dass neben Temperatur und Niederschlag auch Veränderungen der Sonneneinstrahlung einen starken Effekt auf die Gletscher-Schmelze haben können (Huss et al., 2009). Die Strahlung kann über längere Perioden abnehmen, wenn im Mittel verstärkte Bewölkung auftritt, oder wenn Aerosole in der Troposphäre und Stratosphäre einen Teil des Sonnenlichts wieder zurück in den Weltraum reflektieren. Eine solche Strahlungs-Reduktion reduziert die Gletscher-Schmelze *zusätzlich* zu den tieferen Temperaturen, die während solcher Phasen oft vorherrschen. Da es keine Messdaten zur Sonneneinstrahlung gibt, die weiter als bis in die erste Hälfte des 20$^{ten}$ Jahrhunderts zurückreichen, und unser einfaches Berechnungs-Modell die vollen physikalischen Prozesse nicht berücksichtigen kann, beschreiben wir den Effekt von vulkanischen Eruptionen auf die Strahlung mit einem rein empirischen Ansatz. Die Konzentration von vulkanischen Aerosolen wird anhand des Datensatzes bestimmt, welcher Zeitpunkt und Stärke der Eruptionen wiedergibt. In Abhängigkeit der jeweiligen Konzentration fällt jedes Jahr ein Teil der Aerosole aus. Je länger es also keinen (grossen) Vulkanausbruch mehr gibt, desto reiner ist die Atmosphäre und desto mehr Sonnenlicht gelangt auf die Gletscheroberfläche. Die Konstanten der Gleichungen werden so angepasst, dass die beobachteten Änderungen der Gletscher seit 1900 optimal reproduziert werden.

In unserem einfachen Modell beschreiben wir einen weiteren, sekundären Effekt mit stark vereinfachten Gleichungen: Wenn die vulkanischen Aerosole aus der Atmosphäre ausfallen, werden sie zum Teil auf der Gletscheroberfläche abgelagert. Dort entwickeln sie eine gegenteilige Wirkung, indem sie den Schnee und das Eis dunkler machen. Damit wird mehr Strahlung absorbiert und daher die Schmelze verstärkt. Wir berücksichtigen auch, dass die Verunreinigungen mit der Zeit, zum Beispiel durch Schmelzwasser, wieder von der Gletscheroberfläche ausgewaschen werden.

## 4. Resultate

Die Berechnungen zeigen, dass die Veränderungen der Gletscher in den letzten 250 Jahren ungenügend abgebildet werden kann, wenn nur Temperatur und Niederschlag ins Modell einbezogen werden (blaue Linie in Abb. 3). In diesem Fall sagt das Berechnungs-Modell ein fast konstantes Eisvolumen zwischen 1800 und etwa 1940 voraus, was eindeutig nicht der Realität entspricht. Die beobachteten Veränderungen der Gletscherlänge deuten auf einen massiven Vorstoss in verschiedenen Teilen der Alpen zwischen 1810 und 1820 hin, was genau auf den Zeitpunkt der Giétroz-Katastrophe fällt. Nach 1855 zogen sich die Gletscher stark zurück und stiessen um 1890 und 1920 nochmals kurz vor (schwarze Linie in Abb. 3). Mit Temperatur und Niederschlag kann dieses Verhalten nicht erklärt werden. Wenn wir allerdings auch die Effekte von vulkanischen Aerosolen und Verunreinigungen der Gletscheroberfläche miteinbeziehen, kann der generelle Verlauf der Gletscher-Veränderung seit 1800 deutlich besser abgebildet werden (orange Linie in Abb. 3). Dabei zeigt sich, dass vor allem der Effekt der Aerosole auf die Sonneneinstrahlung wichtig ist. Der gewaltige Massengewinn der Gletscher zwischen 1810 und 1820 kann mit diesem Berechnungs-Modell reproduziert werden, wie auch der starke Verlust nach 1850. Die weniger starken Gletscher-Vorstösse um 1890 und 1920 werden ebenfalls vom Modell wiedergegeben. Die Aerosol-Konzentration in der Atmosphäre, das heisst primär der Effekt von vulkanischen Eruptionen auf die Sonneneinstrahlung, spielt offenbar eine absolut zentrale Rolle, um das Verhalten der Gletscher während dem Maximum der Kleinen Eiszeit richtig zu erklären.

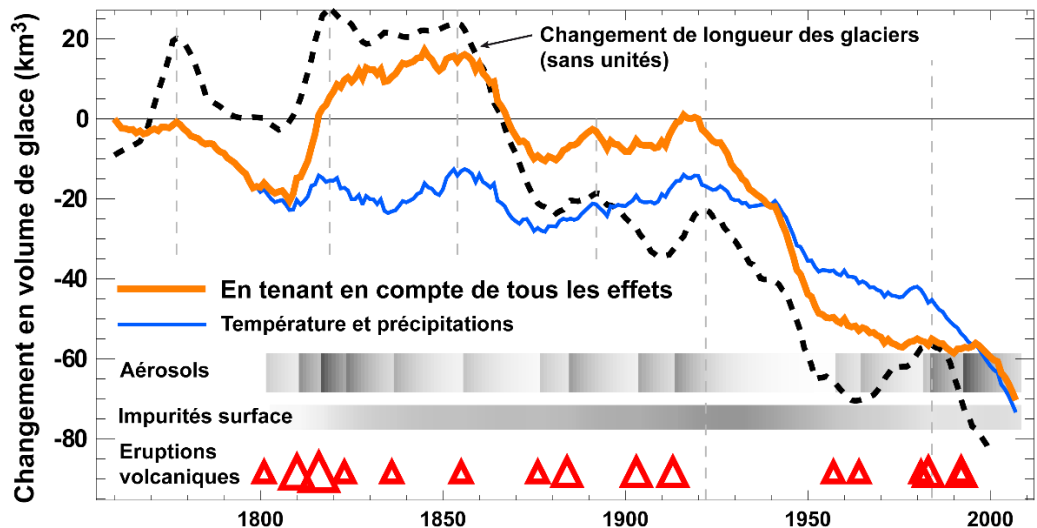

**Abb. 3**: Modellierte Veränderung des Eisvolumens aller Schweizer Gletscher falls nur Schwankungen von Sommer-Temperatur und Niederschlag berücksichtigt werden (blaue Linie), oder daneben auch Effekte von vulkanischen Aerosolen und Verunreinigungen an der Gletscheroberfläche in die Berechnungen einfliessen (orange Linie). Balken zeigen die relative Konzentration von Aerosolen und Verunreinigungen (hohe Konzentration: dunkel). Der Zeitpunkt wichtiger Vulkanausbrüche ist mit roten Dreiecken gekennzeichnet. Die gestrichelte schwarze Linie entspricht der zusammengefassten Änderung der Länge von sechs Gletschern in den Alpen (siehe Abb. 1). Diese Beobachtungen sind ohne Einheit dargestellt und zeigen den tatsächlichen Verlauf der Gletscher-Veränderung seit 1760 auf. Graue vertikale Linien markieren lokale Maxima der beobachteten Gletscherlänge und erlauben den direkten Vergleich zu den berechneten Eisvolumina.

## 5. Schlussfolgerung

Welche Faktoren waren also verantwortlich für den Vorstoss des Glacier du Giétroz, der 1818 zur Bildung des Sees und zur Katastrophe führte? Die Sommer zu dieser Zeit waren kühl, vor allem in den Jahren unmittelbar vor 1818. Allerdings zeigen unsere Berechnungen, dass dies für einen so markanten Gletschervorstoss, der zur Bildung des mächtigen Eiskegels an der Stelle des heutigen Mauvoisin-Staudamms führte, bei Weitem nicht ausreichend ist. Der Effekt der vulkanischen Aerosole, welche zusätzlich zu den tiefen Temperaturen zu einer deutlichen Reduktion der Sonneneinstrahlung führte, muss berücksichtigt werden. Aufgrund von zwei sehr starken Vulkanausbrüchen – einem nicht genau lokalisierten Ereignis 1809 und der Eruption des Tambora im Jahr 1815 (Cole-Dai et al., 2009) – war die Konzentration der Aerosole in den Jahren vor dem Giétro-Ereignis wahrscheinlich tatsächlich so hoch wie sonst nie in geschichtlicher Zeit. Die Kombination von tiefen Temperaturen (unter anderem 1816 im «Jahr ohne Sommer») mit sehr geringer Sonneneinstrahlung hat zu stark reduzierter Schmelze geführt, und damit zu einem zwar relativ kurzen, aber in den letzten Jahrhunderten beispiellosen Massengewinn der Gletscher zwischen 1810 und 1820 geführt. Dieser führte praktisch sofort zu einem starken Vorrücken der Gletscherzungen (Abb. 1). Die relativ hohen Temperaturen zwischen 1850 und 1880, zusammen mit ziemlich trockenen Bedingungen (Abb. 2), bewirkten dann aber einen schnellen Gletscher-Rückgang. Dieser wurde begünstigt durch (1) die tief liegenden Gletscherzungen, die eine starke Schmelze begünstigen, (2) den wahrscheinlichen Anstieg der Sonneneinstrahlung aufgrund relativ weniger Vulkanausbrüche, sowie (3) möglicherweise durch die Ablagerung von Aerosolen auf Schnee und Eis. Unsere Betrachtung zeigt, dass Vorstoss und Rückgang der Alpengletscher während des Maximums der Kleinen Eiszeit ein komplexer Prozess ist und nur durch eine Kombination verschiedener Faktoren erklärt werden kann. Dennoch lassen das stark vereinfachte Modell und die unsichere Datengrundlage ein grosser Spielraum für die Interpretation offen, und genauere Studien sind nötig.

## Referenzen